\documentclass[twocolumn,showpacs,preprintnumbers,aps,prl,superscriptaddress,amsmath,amssymb]{revtex4}

\usepackage{graphicx}
\usepackage{bm}
\newcommand{\srruo}{Sr$_2$RuO$_4$}
\newcommand{\Tc}{\ensuremath{T_c}}
\newcommand{\Ce}{\ensuremath{C_e}}
\newcommand{\Vb}{\ensuremath{V_b}}
\newcommand{\Hcc}{\ensuremath{H_{c2}}}

\newcommand{\vdir}[1]{$\langle #1 \rangle$}
\newcommand{\vect}[1]{\bm{\mathrm{#1}}}

\newcommand{\Fig}[1]{Fig.~\ref{#1}}

\newcommand{\etal}{\emph{et al.}}
\newcommand{\tfullwidth}{8.0cm} 
\newcommand{\twidth}{6.4cm}

\begin{document}

\preprint{PRL/\srruo}

\title{mK-STM Studies of the Temperature- and Field-dependence of the 
Quasiparticle Spectrum of Sr$_2$RuO$_4$}

\author{C. Lupien}
\altaffiliation{Department of Physics, Universit\'e de Sherbrooke,
  Sherbrooke, Qc J1K 2R1, Canada.}
\affiliation{Department of Physics, LASSP, Cornell University, 
             Ithaca NY 14853, USA.}
\author{S. K. Dutta}
\affiliation{Department of Physics, University of Maryland, College Park, 
               MA 20742, USA.}
\author{B. I. Barker}
\affiliation{Laboratory for Physical Sciences, University of Maryland,  
              College Park, MA 20740, USA.}
\author{Y. Maeno}
\affiliation{Department of Physics, Kyoto University, Kyoto 606-8502, JAPAN.}
\author{J. C. Davis}
\email{jcdavis@ccmr.cornell.edu}
\affiliation{Department of Physics, LASSP, Cornell University, 
             Ithaca NY 14853, USA.}


\date{\today}

\begin{abstract}
Atomic-resolution scanning tunneling spectroscopy is used to study the
spectrum of quasiparticle states in superconducting Sr$_2$RuO$_4$. The
measured temperature dependence of the quasiparticle spectrum is
consistent with a multi-band superconducting order parameter with one
band exhibiting a line of
nodes, if a constant density of states offset can be ascribed to a
surface effect.  In magnetic fields $H<H_{c2}$, a vortex lattice, square
and oriented parallel to the $\langle 110 \rangle$ direction of the RuO$_2$
lattice, is detected by spectroscopic imaging. 
Each vortex exhibits one flux quantum and a strong zero-bias conductance peak.

\end{abstract}


\maketitle

\srruo\  displays an exotic form of superconductivity at temperatures
below 1.45 K \cite{maeno94,mackenzie03}. Because of the independence of
the spin susceptibility on temperature below
\Tc\  \cite{ishida98,ishida01,duffy00}, the Cooper pairs are believed to
be in a spin triplet `equal spin pairing' (ESP) state. Muon spin
rotation studies also show a time-reversal-symmetry breaking (spontaneous
magnetization) signal which is also consistent with spin-triplet
superconductivity \cite{luke98}.

\begin{figure}
\includegraphics[width=\tfullwidth]{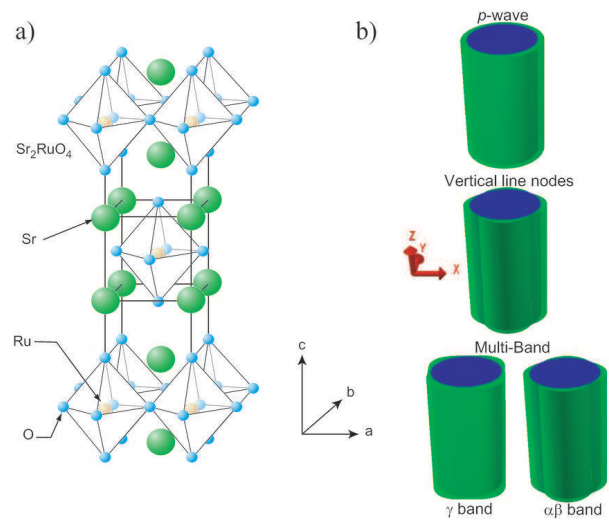}
\caption{\label{fig:structure} a) Crystal structure of \srruo. b)
Schematic of various gap order: top is a $p$-wave, middle is vertical
line nodes and at the bottom is a multi-band model with both a
modulated $p$-wave and vertical line nodes on different Fermi
surfaces. }
\end{figure}

Many discussions have proceeded in terms of a purely
2-dimensional version of the $p$-wave order parameter (OP) $\vect{d} = \Delta
\vect{\hat{z}}(k_x \pm ik_y)$.  But it now appears that the \srruo\ OP
is somewhat more complex than this. It has recently emerged that (i)
the low temperature electronic heat capacity $\Ce$ \cite{nishizaki00}, 
(ii) the nuclear spin relaxation time \cite{ishida00},
(iii) the thermal conductivity \cite{kappa01}, (iv) the London 
penetration depth \cite{bonalde00} and (v) the ultrasound attenuation 
\cite{lupien01} have low temperature power laws which are consistent
with the presence of line nodes (\Fig{fig:structure}b). 

A microscopic model consistent with these observations has also
recently emerged. In \srruo, three electronic bands cross the Fermi
level leading to three closed Fermi-sheets in momentum space: $\alpha$ (X
centered hole-like), $\beta$ and $\gamma$ 
($\Gamma$ centered electron like) \cite{bergemann,damascelli}.
It is proposed that on the $\gamma$ sheet the superconductivity 
is dominant while the remaining two sheets have induced superconducting 
gaps \cite{agterberg97,zhitomirsky01}. Although
this model of \srruo\  is elegant, plausible, and consistent with
experiments, it is by no means certain \cite{mackenzie03}.

Tunneling spectroscopy has historically played a key role in the study
of superconductivity \cite{mcmillan69}. This is because the differential
tunneling conductance $g=dI/dV$ at bias voltage $\Vb$ between a sample
and metal tip is proportional to the sample local density of
quasiparticle states (LDOS) at energy $E=e\Vb$;
$g(\vect{r},\Vb) \propto \text{LDOS}(\vect{r},E=e\Vb)$.  Scanning tunneling
spectroscopic imaging has further advanced the utility of this
technique by allowing mapping throughout space of $g(\vect{r},\Vb)$.
For example, in $s$-wave NbSe$_2$, the vortex lattice and core structure
have been determined with this technique \cite{hess89}, and in $d$-wave
cuprate superconductors it has revealed the vortex structure
\cite{hoffman02}, nanoscale electronic disorder \cite{stmB}, impurity
atom effects \cite{stmC}, and quasiparticle quantum interference
\cite{mcelroy03}.  In this paper we apply these techniques to \srruo.

Several important STM studies of \srruo\ have already been
reported. At temperatures high compared to \Tc, topographic imaging of
surface structure reveals unexpected reconstructions
of SrO surface \cite{matzdorf00}.  Tunneling 
spectroscopy has been carried out below
\Tc\ but on an unidentified, and ramified, surface yielding the
unexpected result that the LDOS at $E=0$ was about 90\% of normal value even
at low temperature \cite{oldstm}. A possible explanation was the level of
disorder seen in the images. Nonetheless, this study was an important
advance since it detected a partially gapped quasiparticle spectrum.

We use a dilution-refrigerator based STM which reaches
15 mK in fields up to 8.5 T with atomic-resolution scanning tunneling
spectroscopy.
This system also allows samples to be cleaved in
cryogenic ultra high vacuum inside the vacuum space of the dilution
refrigerator and then immediately inserted into the STM head mounted
below the mixing chamber.  We use floating-zone-fabricated \srruo\  single
crystals which have \Tc=1.45 K.

\begin{figure}
\includegraphics[width=\tfullwidth]{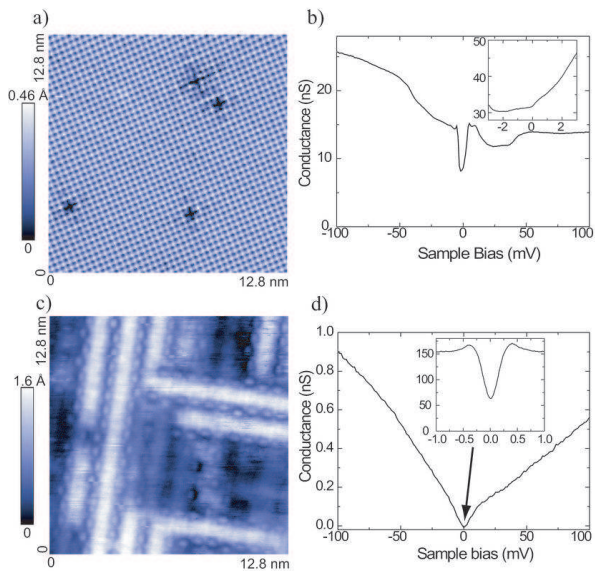}
\caption{\label{fig:topospectra} a) Topographic image of a Ti doped
sample showing the usual square and uniform lattice (and some Ti
impurities). b) Spectra taken on the surface a) away from a Ti atom
showing a 50 mV and a 10 mV gaps. The inset shows that there is no
superconducting feature seen at low energy. c) Topographic image taken
on the pure sample studied here showing a very different surface
structure. d) spectra taken on the image of c) and showing a very
different V-like spectra with weak features (small slope changes) at
50 and 10 mV and reaches a very small conductance around zero
bias. The inset shows data taken at small bias voltage ($\pm$1 mV)
showing the superconducting gap. The topographs were taken with a
setpoint of 100 mV and 0.05 nA.  }
\end{figure}

In \srruo, this cleavage technique can, in principle, result in two
c-axis normal crystalline surface layers: the SrO layer and the
RuO$_2$ layer (see crystal structure in \Fig{fig:structure}a. 
Topographic images of two surface types we observe are shown in
\Fig{fig:topospectra}a and \Fig{fig:topospectra}c. The first surface
layer was identified as being SrO using Ti
atom substitution experiments \cite{barker03}.  This surface never
exhibits a superconducting quasiparticle spectrum but instead shows
the spectrum in \Fig{fig:topospectra}b.

A topographic image of a second type of
cryo-cleave surface (likely RuO$_2$) is shown in \Fig{fig:topospectra}c. 
This image shows a quasi-one-dimensional structure on this
surface.  The width of the ``bands'' in the figure are $4a_0$
(where $a_0$ is the in-plane lattice constant, 0.386 nm) while the
periodicity of the ``beads'' running the length of the ``bands'' is
$2a_0$. Domains of both orientations of the quasi 1-d topographic structure are
clearly seen everywhere on this surface. 
Surface reconstruction effects that are sensitive to 
the cleaving recipe have been seen in some measurements 
\cite{matzdorf00,matzdorf02,shen01}.
It is not clear
how these high temperature cleaved reconstructions are related to the
one reported here.

\begin{figure}
\includegraphics[width=\twidth]{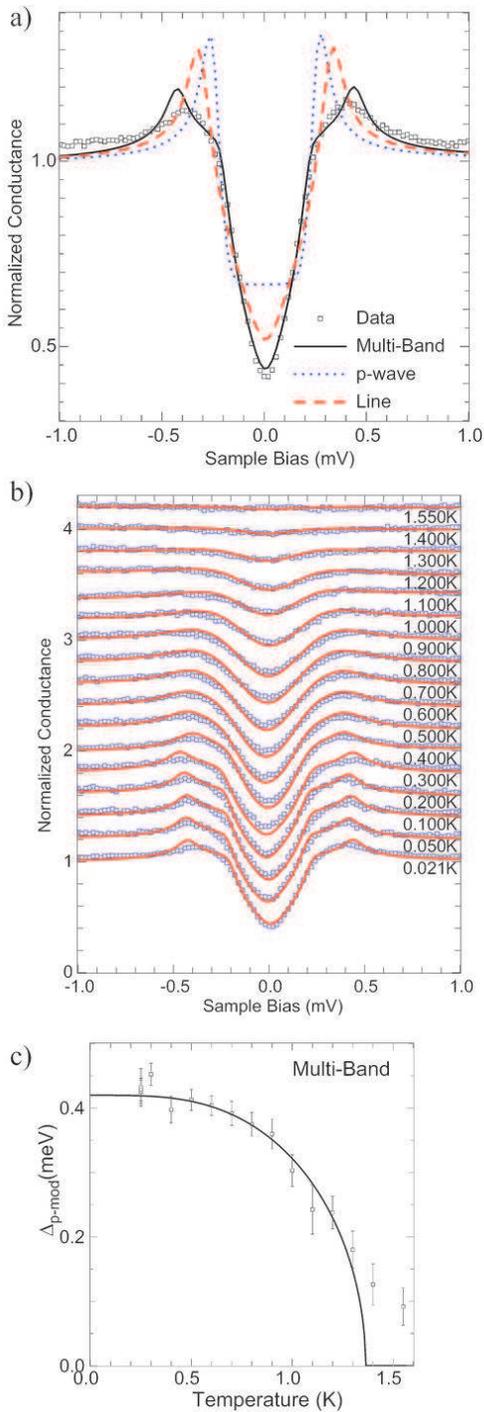}
\caption{\label{fig:temperature} a) Shows three different fits to the
lowest temperature conductance data. The dotted (blue) line is for an isotropic
$p$-wave gap. The dashed (red) line is for vertical line nodes and the
solid (black) line is for the multi-band model discussed in the text.
b) Temperature dependence of the data and the
associated multi-band model fits. The different temperatures are
shifted vertically for clarity.
 c) The $\Delta_\text{p-mod}$ gap
parameter obtained from the fit in b) compared with a BCS like temperature
dependence shown by the line 
($\Delta_\text{line}$ is kept a constant fraction of
$\Delta_\text{p-mod}$ hence has the same temperature dependence).}
\end{figure}

Independent of structural details of this RuO layer reconstruction,
the surface shown in \Fig{fig:topospectra}c displays a differential
tunneling conductance spectrum as shown in \Fig{fig:topospectra}d, and
with high resolution in \Fig{fig:temperature}a).
We have studied the spectrum over regions up to 1.1 $\mu$m$^2$ and it
is highly repeatable and reliable, having almost identical
characteristics independent of exact atomic details of surface
reconstruction.

\begin{figure}
\includegraphics[width=\twidth]{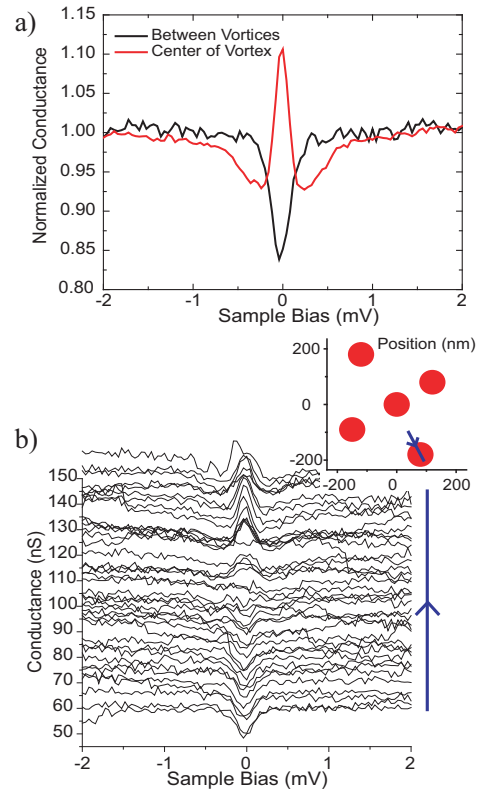}
\caption{\label{fig:field} a) Averaged point spectra at the center
 (red) and in between (black) vortex showing the ZBCP
 at the center. b) Line cut through a vortex (as shown in the
 inset). The inset also shows the location of the 5 places where we
 identified a vortex core. The data was taken with a setpoint of 2 mV
 and 0.15 nA.}
\end{figure}

We carried out the following tests to demonstrate that these spectra
represent the superconducting quasiparticle spectrum of \srruo.
First, the temperature dependence studies show the observed gap
disappears at \Tc\ (see \Fig{fig:temperature}b) filling from 40\% to
100\% of normal $\text{LDOS}(E=0)$.  Second, at millikelvin
temperatures, this gapped spectrum has disappeared at a magnetic field
of $\mu_0 H=100$ mT, which is very close to the c-axis $\Hcc$ of the
superconducting state. Third, in the presence of fields $H<\Hcc$
this spectrum shows the characteristic spatial variations expected
from the core of vortices exhibiting $\phi_0$ of magnetic flux (see
\Fig{fig:field}). We therefore conclude that we are dealing primarily
with the quasiparticle spectrum of superconducting OP of \srruo.

To analyze the shape of the spectra and their temperature dependence
we consider three
different models of OP; an isotropic $p$-wave gap 
\begin{subequations}
\label{eq:allgaps}
\begin{equation}
\label{eq:pgap}
  \Delta(\phi) = \Delta^p_0 e^{\pm i\phi},
\end{equation}
a gap with a vertical line of nodes 
\begin{equation}
\label{eq:vline}
  \Delta(\phi) = \Delta^\text{line}_0 e^{\pm i\phi}\sin(2\phi),
\end{equation}
and a modulated $p$-wave gap (which is combined with vertical line
nodes for the multi-band model): 
\begin{equation}
\label{eq:pmod}
  \Delta(\phi) = \Delta^\text{p-mod}_0 e^{\pm i\phi}
               \left [ 1-\frac{\alpha}{2}- \frac{\alpha}{2}
               \cos(4 \phi) \right].
\end{equation}
\end{subequations}
In this latter case,
the in-plane modulation of the gap along $\phi$ has
peak-to-peak amplitude of $\alpha$ (i.e. a maximum of 1 and a minimum
of $1-\alpha$). 

The local conductance $g(\vect{r}, \Vb, T)$ is obtained from the local
density of states of the various gap models (Eq. 1a-1c) by including thermal
smearing $T$. Therefore predictions for the 
differential tunnel conductance of the gaps are obtained as $g_p(V)$ 
(which could as well represent an $s$-wave OP), $g_\text{line}(V)$ and
$g_\text{p-mod}(V)$.
Note that horizontal line nodes ($\Delta_0 \cos(k_z c)$) gives the 
same conductance as
vertical line nodes, and that the orientation of the nodes as well as
the $\exp({\pm i\phi})$ part of the gaps also does not modify the conductance.

To relate the three models to the measured conductance $g(\vect{r},\Vb)$ in the
experiment (\Fig{fig:temperature}a), we fit the data to a general form 
which includes a background density of states which is added as a 
fraction $f$ of the total conductance,
\begin{equation}
\label{eq:gfit}
  g(V)=A\Bigl[  \left( 1-f\right) G(V)   + f     \Bigr]
\end{equation}
and for the multi-band model
\begin{equation}
\label{eq:multiband}
  g(V)=A\Bigl\{  \left( 1-f\right) \left[ 
                   \left(1-x\right) g_\text{line}(V) + 
                    x g_\text{p-mod}(V) \right]  + f     \Bigr\}
\end{equation}
where $V$ is the bias voltage, $A$ is the amplitude of the
conductance and $G$ is either $g_\text{line}$ or  $g_p$. We consider
$f$, which is about 40\%, to be a spurious 
effect. This is reasonable since heat
capacity experiments show the number of thermally occupied states in bulk 
falling precipitously at these low temperatures. 
In case of
the multi-band model, $x$ is the fraction of the non-background LDOS due
to the modulated $p$-wave gap (the $\gamma$ band). Therefore we
have 2 or 3 adjustable parameters ($A$,  $f$ and $x$) in addition to the
predicted $g(V)$ of the gap function of interest. The temperature
is fixed to the measured sample temperature, except at
very low temperatures below 250 mK where we fix the temperature to 250
mK because (1) we believe from independent studies that this tip did
not cool below that temperature and (2) this is the sample temperature
at which the measured spectra start changing.

As can be seen in \Fig{fig:temperature}a, the simplest $p$-wave fit
$G=g_p$ is quite poor compared to the line nodes. The multi-band model
is better than a pure line nodes model (possibly because of
the larger number of parameters). Other gap-models with extra
parameters over a simple line node also fit as well, such as a modulated
horizontal line of nodes: $\Delta_0\cos(k_z c)\left[1-\beta \cos(4
\phi)\right]$ (but this would fail to reproduce the heat capacity jump
at \Tc).  The particular multi-band model was selected because
it is consistent with a recent study of the dependence of the heat
capacity on the angle of the magnetic field \cite{deguchi04}. 
We caution that the value of $x$ is controlled not only by the
density of states of the different bands but also by the tunneling
matrix element.

We next study the temperature dependence of these tunneling
spectra. In \Fig{fig:temperature}b, we show averaged spectra taken at the 
17 temperatures between 20 mK and 1.6 K. They were obtained after
averaging about 100 individual spectra (25 s each) with a bias
modulation of 20 $\mu$V$_\text{RMS}$ and setpoints of 10 mV / 0.25 nA
or 5 mV / 0.15 nA. Different locations were used for data acquisition at each
temperature.
The solid lines in \Fig{fig:temperature}b are the result of fits to
the multi-band  model.  The lowest temperature is fitted the
same way as in \Fig{fig:temperature}a above. For all other temperatures 
$f$, $x$ and $\alpha$ are kept constant at their low temperature
value (0.38, 0.30 and 0.37 respectively), $T$ is set to
the measurement temperature but $\Delta^\text{p-mod}_0$ is left to evolve while
keeping the ratio $\Delta^\text{p-mod}_0/\Delta^\text{line}_0$ fixed to the
low temperature value of 1.9.

From all these fits, a temperature dependence of the gap 
$\Delta_0^\text{p-mod}$ can be extracted and the result is shown in 
\Fig{fig:temperature}c where the solid curve is a fit to a BCS
temperature dependence of the gap. A very similar graph is
obtained if a vertical line nodes model is used. 

A second important use of spectroscopic mapping is
in the determination of effects of magnetic field on the
superconducting OP. This can be very revealing 
since the structure of the vortex lattice and
the electronic structure of the core are closely related to the
symmetry of the OP. For the case of a $p$-wave OP, a number
of theoretical studies \cite{tewordt01,takigawa01,maki99,ichioka02} have 
explored the predicted structure
of the vortex core in 2-d OP $\vect{d}=\Delta\vect{\hat{z}}(k_x\pm ik_y)$.

Experimentally, the flux lattice has been measured by small angle neutron
scattering \cite{riseman98}. 
However, until the present work, nothing is known about the local
electronic structure of \srruo\  vortex cores. To explore these 
issues we mapped  point spectra 
at $>$140 individual locations in  $B<\mu_0 \Hcc$. 
We detected 5 vortices with mean separation 
between cores of 180$\pm$30 nm. Within our statistical uncertainty, the
lattice is square and oriented parallel to the \vdir{110} axes of the crystal
lattice which is quite consistent with previous neutron studies. 
We detect the vortices from the changes in $g(\vect{r},0)$ at the
vortex center, forming a pattern consistent with a square lattice and
the separation of the core locations at $B=50$ mT yields a flux $\Phi=1\pm0.15
\Phi_0$ threading each core.

To study the \srruo\  vortex core itself, we measure $g(\vect{r},V)$ 
at regions nearby
the core (\Fig{fig:field}a) and along lines connecting two cores 
(crystallographic
\vdir{110} direction) as shown in \Fig{fig:field}b. The most striking 
feature is the
very strong zero bias conductance peak (ZBCP) of FWHM ~250 $\mu$eV and
maximum intensity 115\% of the normal density of states. The coherence
length can also be estimated from decay length of the gap suppression
away form the core to be 50 nm in reasonable agreement with the
estimate of 65 nm from penetration depth studies.

A number of important
conclusions can be drawn. First, the RuO surface of \srruo\  exhibits a
gapped, particle-hole symmetric, density of states spectrum consistent
with a superconductor. 
Second, the temperature dependence of the observed $g(\vect{r},E)$ is
best fit by a superconducting multi-band gap with at least one band supporting
a line of nodes \cite{pmod}. This
provides further corroboration for this picture of the gap structure
in \srruo . Third, direct imaging of the vortex lattice by
spectroscopic mapping is now possible. Because of this, the long
mooted experiment to search for chiral domains that can exist in a
$k_x\pm ik_y$ OP \cite{dumont02}, becomes
technically feasible.  Finally, the large ZBCP discovered in \srruo\  is
consistent with $p$-wave models that did not consider line nodes. 
It will require new theoretical models to understand the ZBCP in a
multi band model.

\begin{acknowledgments}
This work was supported by the ONR grant N00014-03-1-0674, NSF grant
NSF-ITR FDP-0205641, ARO grant DAAD19-02-1-0043 and MEXT of Japan. 
C.L. aknowledges support from NSERC. 
\end{acknowledgments}


\end{document}